# Emerging Topics in Assistive Reading Technology: From Presentation to Content Accessibility

**Cynthia Chen**

Lexington High School, Lexington, MA
lexcynthiayc@gmail.com

**Peter Fay**

IBM Accessibility Research

Kendall Square Cambridge

peter_fay@us.ibm.com

**Abstract**

With the recent focus in the accessibility field, researchers from academia and industry have been very active in developing innovative techniques and tools for assistive technology. Especially with handheld devices getting ever powerful and being able to recognize the user's voice, screen magnification for individuals with low-vision, and eye tracking devices used in studies with individuals with physical and intellectual disabilities, the science field is quickly adapting and creating conclusions as well as products to help. In this paper, we will focus on new technology and tools to help make reading easier--including reformatting document presentation (for people with physical vision impairments) and text simplification to make information itself easier to interpret (for people with intellectual disabilities). A real-world case study is reported based on our experience to make documents more accessible.

## 1. Introduction

For individuals who are classified with having intellectual disabilities, they are usually diagnosed before age 18, during the developmental period. With IQ's usually between 70-75, they may have problems socially, conceptually, and lacking practical skills, such as keeping track of time (*Definition of Intellectual Disability*, 2017). Individuals with cognitive disabilities function on a different plane, as they may have difficulty remembering things or paying attention during social or academic settings. However, the amount of intelligence individuals with cognitive disabilities is not affected by their disability, as many individuals, such as Albert Einstein, was categorized on the spectrum as having dyslexia (*Cognitive*, 2013).

The push for technology to be accessible for all individuals did not gain much traction until 1999, when the World Wide Web Consortium released the WCAG 1.0 (Web Content Accessibility Guidelines), which recommends techniques that web developers can implement to make their websites and other applications accessible for individuals with disabilities. The release of WCAG 2.0 in 2008, which became an ISO (International Organization for Standardization) in 2012. WCAG 1.0 highlighted 14 guidelines, with ideas such as creating clear and simple interfaces, allowing for both visual pictures and read-out-loud text as options, and making sure that instructions are easy to follow. Within each guideline was 65 checkpoints, with priorities ranking 1, 2, and 3 (A, where web developers must meet the standard, AA, where web developers should meet the standard, and AAA, where web developers may meet the standard). WCAG 2.0 elaborates more on guidelines and checkpoints, with additions including: alt texts, where pictures or visuals have a short text describing what the depiction is, and tab order, where users can tab logically to the different parts of a presentation (Henry, 2017).

WCAG guidelines have been adopted globally, as businesses in the United Kingdom and United States can be sued if they do not provide accessible websites for their users (Hensley, 2015). Particularly in the United States, the push for equal access for all web users has been progressive, as there are also ethical obligations. Section 508 of the Rehabilitation Act of 1973 adopts seventeen of the WCAG 2.0 guidelines (*A Guide to Disability Rights Laws,* 2009). In the accessibility field, researchers have developed various tools to see how individuals with disabilities are interacting to the web.

One method that has led to many breakthroughs in web access is eye tracking. What most individuals do not realize is that websites are often times very distracting, with ads popping up in various locations, videos auto-playing, and font size and colors being unsuitable for individuals who are color-blind or are unable to focus on text. Eye tracking is not a new form of research, as it has been around since the 19th Century, but it has gained a lot more popularity in the 21st Century, with researchers recruiting both individuals without disabilities and those who have disabilities, to see how their eyes function on a sample website (Leggett, 2010). This method is useful as the eye tracking device is able to track the pupil movement to see what it focuses on first, what is distracting (i.e. causes them to shift their attention to the distraction), and how long it takes for an individual to finish reading a paragraph or looking at a visual, to help organizations make accessible adjustments to their websites.

Web accessibility is especially important as the modern world has shifted to become more environmentally friendly, with schools and companies transforming paperless, so the need for accessible technology is crucial. These companies have created accessibility teams within their areas of focus, as seen through Apple's Siri function, which allows users to use their voice to perform daily tasks, such as scheduling doctor appointments, checking the weather, and making a phone call. IBM is focusing their attention on Watson Artificial Intelligence platform, which is enabling a relationship between computers and humans. For example, computers are able to recognize a "cat" not because it has "seen" one before, but it uses machine learning software to train the computer to recognize a cat and then evaluate its confidence level that the image that it's analyzing is likely to be a cat.

This paper is organized as follows. In section 2 we will present a few software tools to re-present information in a more accessible format/media (e.g., text->text, text->voice, image->text), and a real-world case study is included based on our experience in making documents more accessible. In section 3 we will discuss an emerging technology that is very useful for assistive reading technology: eye tracking. To help individuals with intellectual disabilities, text reformatting is not enough, and an emerging research topic called text simplification is presented in section 4, which aims to simplify the content of text and make it easier to interpret. We conclude in section 5.

## 2. Accessibility Through Presentation and Formatting

In this section, some popular technology and tools will be discussed to show how to improve accessibility in reading through careful design in text presentation and formatting.

**Microsoft Office**

Microsoft Office software has been very accommodating for web developers and users, in the sense that they have created an accessibility tool that allows for an instant "Accessibility Checker" which highlights issues within a Word document, PDF, PowerPoint presentation, etc. Users can then create alt-texts for diagrams, pictures, and visuals, as well as check the reading order, color contrast, table organization, and much more. Word then gives the user tips and warnings on their document. To check the color contrast for individuals who may be color blind or low vision, users can opt to look at the Word document in gray-scale mode, where colors are turned to various shades of black, gray, and white. The built in MS Office accessibility tools allow users to manually resolve accessibility issues on documents; it also provides a foundation of knowledge on how to make your content accessible. (*Use the Accessibility Checker on Your Windows Desktop to Find Accessibility Issues*, 2016).

The W3C has developed a comprehensive set of accessibility guidelines known as checklists. During this research project, we delved more deeply into the accessibility checklist and learned that there were certain guidelines that application and content developers must follow, if they want to deliver an accessible solution. Alternative text, or Alt text for short, must be applied to every single diagram, visual, or picture in document/presentation or on a web page. This ensures that users with low or no vision can use the text-to-speech feature to hear what the graphic was depicting. Unacceptable color pairings (red/green, neon colors, light blue on dark blue, etc.) was changed to either yellow text on a black background, or with sufficient contrast to make sure that the content was easy to read. (*Make Your PowerPoint Presentation Accessible*, 2016).

Powerpoint slides that were too "busy", in the sense that the document was too cluttered were simplified. There was a set order to every document that needed to be made accessible, as there was a specific font type (Times New Roman) and font size (no smaller than size 14 for text). One of the fundamental guidelines was that the reading order, or tab order, needed to be in a logical order so that a blind user accessing the content with a screen reader would be able to easily navigate the content. If the reading order was incorrect, the blind user would end up hopping around the page--from the title, to a visual caption and then back to the text itself. This scenario would be very confusing for the user and make it much harder to comprehend the information in the document. For particularly detailed visuals, it is difficult to explain the meaning just using the alt text. In those cases, we added we added a notation at the end of the alt-text guiding the user to refer to the speaker notes for more information. The speaker notes provided lecturer's notes and often times explained in detail what the visual was representing. Hyperlinks embedded within each document needed to be checked to make sure that they worked and were not outdated (*Make Your PowerPoint Presentation Accessible*, 2016).

Word documents and PowerPoint presentations are relatively easier to make accessible than Adobe PDF's. Microsoft provides integrated accessibility checking tools built right into their products. With PDF's, Adobe Acrobat Pro must be used to scan the PDF to find the accessibility violations. Every alt text, speaker note, color changes, and other accessibility errors must be corrected and saved in the new PDF file, in order to make it accessible (*Create and Verify PDF Accessibility (Acrobat Pro)*, 2016).

**Apple's Voiceover and Siri**

Apple introduced its VoiceOver option for computers in 2005, where users could perform daily tasks using only their voice. The company quickly revealed more accessible technology, including Braille display, creating devices that were completely accessible to the blind, and adding VoiceOver to iPhone, iPods, and TV's. Every iPhone in the status quo comes equipped fully with accessibility features that can easily be activated with a quick visit to the Settings app (*Blind Faith: A Decade of Apple Accessibility*, 2011).

**IBM Watson**

IBM is taking a different approach from Apple, who is primarily focused on creating hands free technology and software. With IBM Watson, its main focus is on coding software that is able to recognize images and break down its properties and narrow down the identity of the image to several options. Computers are different from humans in the sense that if a human sees a tree, they know it is a tree even if all trees are different, because the brain can deduce that it has seen a similar image, which is that of a tree. Computers are unable to do this, as they know it is a tree because it takes them a significant number of pictures of trees to recognize that the new image is a tree, and even then, it is still not 100% certain that it is a tree.

Specifically, within the Aging and Accessibility team at IBM Watson in Cambridge, MA, the focus is on helping the elderly with daily tasks, and creating programs that allow their family to interact with them in case there is an emergency. For example, if an elderly individual has been going to the bathroom for an excessive amount of time at odd hours, it is probably the case that there is something unusual happening, which the application would then send an alert to an immediate family member to notify them and let them know what was happening. The family member could then message to the application to find out what happened. The monitoring application uses artificial intelligence (AI) to deduce the events that led to the emergency. If the elderly individual has been sleeping significantly longer than what is considered normal for them, the family member could also receive questions from the AI system asking them what they ate and if they were experiencing anything unusual.

**A Real-World Case Study**

Based on the internship work during Summer 2017 at IBM, using Microsoft PowerPoint as an example, specific changes are made to the slides to accommodate individuals with low vision and other vision impairments. Here are some of the best practices for making slides accessible that we learned during the project:

1. Alt text pictures

    Alt text, otherwise known as alternative text, is a short text description of every graphic. This helps users who have low to no vision and are using screen readers to be able to interpret the graphic. This is a necessary checklist on accessible documents.

2. Graphic simplification

    Graphics that have many components are screen grabbed and replaced with one graphic that contains all of the components. This is because creating alt texts for every minute detail is not only time-consuming, it is also confusing. Having one or two graphics per slide allows readers to follow along the presenter with ease.

3. Background simplification

    An issue with some PowerPoint slides is that their backgrounds can be rather confusing. Whether there is too much color in the background or patterns come into play, the key to accessible slides is to keep everything as simple as possible. Refrain from using patterns, cute puppies, and other graphics as backgrounds, as it can take away the attention from the text of the slides.

4. Reading order

    Reading order is a key component for accessible documents, as when the reader is tabbing through the PowerPoint, if something is out of order, it can cause confusion. By going to the formatting pane and manually moving the order of the title, text, graphics, caption, etc., it allows the reader tabbing through the content or using a screen reader to follow the content on the slides in a logical order.

5. Font size and texts

    For every PowerPoint, project titles needed to be size 54, slide titles size 32, texts were size 20, and captions were no less than size 14. This allows for more control over the formatting of slides. If the text is too small, readers are unable to read it. It's better to have a standardized format for all slides, as it creates more order.

## 3. Reading Comprehension Research Using Eye Tracking Technology

Eye tracking was first utilized by Edward Huey in 1908, where he created a device that allowed him to see when the individual's eyes drifted away from the text (Huey, 1908). Although he did not have the modern technology tools that would have allowed him to track the eye movement, such as MRI or fMRI techniques, he acknowledged in his book, *The Psychology and Pedagogy of Reading*, that he thought tracking eye movement precisely would be impossible (Huey, 1908). 70 years later, eye tracking gains some popularity, as advertising companies began utilizing this tool to see what consumers spend more time reading, what they skimmed, and what they completely ignored (Leggett, 2010).

Towards the end of 1990, web designers started to focus on creating better websites. Eye tracking was used to see what the web user would focus their attention on, as websites generally had the same design as newspapers (Leggett, 2010). Nowadays, tech companies are looking to encompass all the data that they collected using eye tracking software to change website interfaces, how they relay information, and the best method to showcase their products so that individuals with low-vision, prone to seizures, color blindness, and other limitations can access their sites with ease.

There are many students that have been diagnosed with having a learning disability in school, such as dyslexia and autism. Eye tracking studies that are focused on children are especially useful, as disabilities are unique across the board, and with data that is specific for each age group, researchers are able to access more accurate data.

The importance of eye tracking technology is that it allows researchers to record which words individuals have trouble with and figure out the reason why they are having issues with comprehension. Whether it was phonetically problematic or the way that the letters were arranged in the words was difficult, text simplification was the next step. This topic we will delve more into in the next section.

**Children with Dyslexia**

```
       n."saidB    y. "W              r    edon'    nother
"Comeo    ets    ehav   pi cku  i  o n.W    thavea
                 eto        pth   sc
        fpopc                              loor?"
  cano    corn.""Arew   ngt   t popcor  t'sbe   nthef
                         egoi   oea      ntha   eno
 "Wec    wa     Betsyan     ed.
    an    shit."           swer
 "Tha    goodi       an."Wec    wa      meo    lofy     puspi
   t'sa    dea,"saidSus    an      shit.Co  n,al   ou.Hel   ck itup."

 Thech    enw   entt      r k.Itt
         ildr       owo    ook    emal      met    kup al  hepo    orn.
                                 th    ongti    opic     lt     pc
 Th    the         mintot
   en    ytookt hec      hek       dBe    ashedi     hec
            o           itchenan    tsyw       t. Allt    hildren
 th    tth    hatw    stt    ingt    Betsy  pu  cornin  obigpans
   ough   att       asju    heth    odo.       the        tw
 toputint
          heoven.
```

*Figure 1 This image shows how an individual with dyslexia comprehends a paragraph of text (MacMillan, 1965).*

Dyslexia, a learning disability, does not affect an individual's intellect, but rather their ability to read and comprehend text. Often times children would be at a lower reading level than their fellow peers because the frustration accompanied with the difficulty of word comprehension would cause the child to give up.

Even for individuals who do not have a reading impairment, it is rather difficult to comprehend the above text. Words seem to be jumping out of order, with letters being jumbled within the words, as well. With all these factors, the reader's comprehension of the words along with their meanings, takes a longer time. The correct reading order is, as follows:

> """Come on." said Betsy. "We have to pick up this corn. We don't have another can of popcorn." "Are we going to eat popcorn that's been on the floor?" "We can wash it." Betsy answered." "That's a good idea," said Susan. "We can wash it. Come on, all of you. Help us pick it up." The children went to work. It took them a long time to pick up all the popcorn. Then they took the corn into the kitchen and Betsy washed it. All the children thought that that was just the thing to do. Betsy put the corn in two big pans to put in the oven." (*Reading Difficulty Simulation,* 2014).

An eye tracking study with dyslexic children has shown that word reading ability was lower than non-dyslexic children. However, auditory word identification was not much different for both groups (dyslexic and non-dyslexic). Their main focus was on phonological rhyme relationships, as the results indicated that there was a clear difference between dyslexic children and non-dyslexic children in the way that they processed the relationships. However, tests that were just for phonological awareness showed conflicting results with eye tracking results. In the phonological awareness tests, dyslexic children did not show significant difference in their results from non-dyslexic children, while in the eye tracking tests, they performed below the control group (non-dyslexic children) (Desroches, Joanisse, Robertson, 2005). This shows how eye tracking offers a more precise measurement of how well an individual performs in a task.

**Individuals with Autism**

An eye tracking study conducted by Pelphrey et al in 2002 showed that individuals who were categorized on the spectrum as having autism spent less time looking at the main features of the face (eyes and nose) when shown a picture of a person, but rather paid more attention to the surroundings of the person and other features, such as their ears, hair, and chin. While the control group, individuals who were not on the spectrum, were geared up with the eye tracking device, they focused more on the eyes and nose when compared to individuals who had autism. They also found that the ability to tell the "fear" emotion on a person's face proved difficult for individuals with autism (Pelphrey et al, 2002). Eye tracking was able to collect all of this information as when an individual focused its attention on a part of the visual, it collected data as to how long their eyes looked at a specific part.

When trials were done for individuals with autism's family members, researchers found that not only were individuals with autism not focusing their attention on another person's eyes, but that their family members were as well. This helped create a factor for diagnosing autism, as the aversion of eyes could be an early indicator that a person has autism (Dalton et al, 2006). Another study that used eye tracking software found that the control group focused two times more on the eye area than individuals who have autism, as the second group focused two times more on the mouth region and surroundings (Klin, Jones, Schultz, 2002).

In next section, we will discuss how to make textual information more accessible (e.g., easier to understand semantically) through text simplification.

# 4. Improving Reading Comprehension Through Text Simplification

Text simplification (TS) is the next step after the eye tracking studies, as it takes the identified difficult words or text pieces and simplifies them to allow for individuals who have disabilities to comprehend the text without altering its original meaning. TS can also be helpful for non-native language speakers, small children, and the elderly to help them understand text on a deeper comprehension level. Examples such as replacing "plethora" with "a large number of things", and "benevolent" with "kind" can go a long way to help readers comprehend the content.

Splicing and reordering are two methods for text simplification. Splicing words, such as "two rope parts" to "parts of a rope", can help comprehension. One can also reorder confusing or complex sentences, such as "the dog-walking man abruptly turned around" to "the man who was walking a dog turned around abruptly".

**Individuals with Intellectual Disabilities**

With one in every ten individuals having an intellectual disability (ID), it is crucial that accessible text be created, as many websites do not currently offer any simplification of text. The majority of individuals with intellectual disabilities have mild forms of this impairment and can typically achieve a middle school reading level. Individuals with ID may also have trouble with living independently and making decisions, as they may not be able to comprehend legal jargon when signing a contract, or fully consent when going to a doctor. When individuals with ID want to apply for a job, they often can not understand the application process or the information employers are seeking. In today's highly connected online world, users need access to information and communication technologies (ICT) to participate fully in life and society. Today online access is typically required when applying for a job or shopping for necessities, and other daily tasks. With the majority of the Fortune 500 companies mandating that job applications be submitted online, it is especially important that text simplification be used, so that individuals with ID can use modern technology (Chen, Rochford, Kennedy, Djamasbi, Fay, Scott 2016).

Individuals with ID have issues comprehending, inferring, and remembering information from text. A study which used three different types of questions, which were either true/false, multiple choice, or multiple choice with visuals attached to the questions, were given to individuals with ID. The researchers wanted to test which of the three options would generate the most correct answers. The results showed that individuals were able to answer simple questions more often when compared to their complex counterparts. However, the data also showed that for questions with visuals attached to them, "ClipArt", the accuracy for simple questions was slightly higher than multiple choice questions. If the answer choices contained numbers, questions that contained visuals or were multiple choice were not ideal. The researchers concluded that even though the addition of visuals may seem distracting, it was overall helpful, as it helped keep the test group engaged within the activity and not lose interest (Huenerfauth, Feng, Elhadad, 2009).

**Manual Text Simplification with Operationalized Plain Language Rules (OPLR)**

Plain English work dates back many decades, as various forms of dictionaries are updated every few years to keep up with the English language (Cutts, PLAIN). We have gone through such work to create the following ten rules. These rules can be used in manual text simplification to produce easier comprehending text.

**Rule 1: Keep it SSS (short simple syllables)**. Words that have multiple syllables should be replaced with their commonly used, shorter syllabled counterparts. Utilize more frequently used words. A good way to check frequency is at Google's own software called Ngram, which can be found here: https://books.google.com/ngrams/graph?content=&year_start=1800&year_end=2000&corpus=15&smoothing=3&share=&direct_url=.

Example: allocate → assign,
copious → abundant

However, credence should be given towards common usage, as multi-syllabled words, such as "yesterday" or "tomorrow" are part of the common English dictionary. Therefore, in a scenario where a web developer is weighing between number of syllables or common usage, they should always pick common usage. A free online tool can be found at www.thesaurus.com with the checkbox for "common" checked, to see complex words and their synonyms.

**Rule 2: Short sentences are often better.** The average sentence length should be no longer than 10 words, any more should be divided into two separate sentences.

Example: "The brown dog was being walked by his owner towards a park, where he would play with other dogs for five hours." Change this sentence to the following: "The owner was walking his brown dog towards a park. The dog would then play with other dogs for hours."

**Rule 3: Don't simplify terms.** Any time the text uses an abbreviation or acronym, be sure to fully write out the entire meaning and then the abbreviation or acronym in parentheses after. When using it later, it is acceptable to use only the abbreviation or acronym. Even if it is a commonly used abbreviation or acronym, such as "PO", be sure to always write "post office (PO)".

Example: "The University of California Berkeley (UCB) has one of the lowest acceptance rates nationwide. UCB is one of the more competitive colleges in California."

**Rule 4: Utilize active voice**, not passive when talking about the present.

Example: "The ice cream was licked by the child." This sentence uses a passive voice. Restructure this sentence to "The child licked his ice cream."

Always keep sentences as concise as possible. The main idea cannot be muddled when simplifying text, so make sure that the central points are not lost within translation.

**Rule 5: Make sure grammar and spelling are always correct**. Spelling checkers, or autocorrect, are great ways to make sure that there are no typos in a text. Microsoft Word has a built-in spell checker, but there are also free online programs such as https://www.grammarly.com/, which spell check any text entered into the website.

**Rule 6: Replace proper nouns with pronouns**, such as "you", "he", "she", etc. for the reader.

Example: "The Patriots fan, Gisele Bundchen, stood up and cheered loudly for her favorite quarterback, Tom Brady." This sentence has many pronouns, which should be replaced, as follows: "The sports fan stood up and cheered loudly for her favorite quarterback."

**Rule 7: ALL CAPS IS NEVER A GOOD IDEA FOR EMPHASIZING IDEAS**. For emphasis, web developers should utilize either bolding or italicizing the text.

Example: "The man jumped up and yelled: "I SHOULD HAVE NEVER BOUGHT ICE CREAM ON A 110 DEGREE DAY!"". This sentence should be restructured to "The man jumped up and yelled: "I should have **never** bought ice cream on a *110-degree* day!""

**Rule 8: "Wassup, dawg?" and other forms of slang, jargon, colloquialisms should never appear in texts**.

Example: "u r so gr8 for buying us tix and fud at the theater! tysm!" This incoherent texting lingo should be changed to the following: "You are so great for buying us tickets and food at the movie theater! Thank you so much!"

Do not remove a word for a shorter message, as it can alter the meaning of the original phrase. In the above example, the original text did not include "movie theater", just "theater". In this instance, it may not seem like a big deal, as individuals can infer that the original sender meant to say "movie theater", but it is better to include the whole phrase.

**Rule 9: Use lists and tables for organization**. When writing about results, data is often included. However, in a giant block of text, it can be hard to follow.

Example: "Here are the directions to my house. Turn right on Willow Street and drive 0.4 miles until you reach I-94. After two red lights, turn left until you reach Boulevard Avenue. Drive 12 more miles until you reach 1 Broadway Street, which is where my house is". Change this paragraph of text to the following: "Here are the directions to my house. 1. Turn right on Willow Street and drive 0.4 miles until you reach I-94. 2. Turn left after two red lights until you reach Boulevard Avenue. 3. Drive 12 more miles until you reach 1 Broadway Street, which is where my house is."

**Rule 10: No double negatives**. Avoid using phrases such as "I don't know nobody…" They can be confusing and hard to understand.

Example: "I don't know nothing!" Besides this commonly used phrase to signify that a person does not know what the question is asking for, it actually means that there is nothing that they do not know, i.e. they know something.

This is already confusing enough for individuals, double negative statements should not be utilized.

**Automatic Text Simplification**

Automatic Text Simplification (TS) belongs to the field of Natural Language Processing (NLP) and can be performed at lexical, syntactic, or discourse simplification levels (Biran, Brody, Elhadad, 2011). Some methods used handcrafted rules, and others use machine learning techniques (Huenerfauth, Feng, and Elhadad, 2009) (Rello, Baeza-Yates, Dempere-Marco, and Saggion, 2013). This line of work has attracted many researchers from both academia and industry. A recent tool is provided at http://158.121.178.171/contribute/ to perform lexical simplifications. Another available online tool is at http://contentclarifier.mybluemix.net/#api, which is created by IBM Accessibility Research for the purpose of text simplification. The IBM AbilityLab Content Clarifier API allows its developers to simplify, summarize, and augment all types of text including: social media, web sites, documents, email, chat messages, etc. By filtering out long and complex sentences and replacing it with a simplified version, users will have an easier time understanding the main idea of the simplified content.

## 5. Conclusion

In this paper we discussed a few new emerging technology and tools to make reading more accessible for people with physical/vision impairments and intellectual disabilities, which includes text reformatting, eye tracking technology, and text simplification. We also reported some of our own work on accessible document presentation and plain language rules for text simplification.